\documentstyle[epsfig]{aipproc}

\begin{document}

\title{Testing the Invariance of Cooling Rate in Gamma-Ray Burst Pulses}

\author{A. Crider$^*$, E.P. Liang$^*$, and R.D. Preece$^{\dagger}$}
\address{$^*$Rice University, Houston, TX 77005-1892\\
$^{\dagger}$University of Alabama at Huntsville, Huntsville, AL 35899}

\maketitle

\begin{abstract}
Recent studies have found that the spectral evolution of pulses within 
gamma-ray bursts (GRBs) is consistent with simple radiative cooling.
Perhaps more interesting was a report that some bursts may
have a single cooling rate for the multiple pulses that occur within it.
We determine the probability
that the observed ``cooling rate invariance'' is purely coincidental
by sampling values from the observed distribution of cooling rates.
We find a $0.1-26\%$ probability that we would randomly
observe a similar degree of invariance based on a variety of pulse
selection methods and pulse comparison statistics.
This probability is sufficiently high to
warrant skepticism of any intrinsic invariance in the cooling rate.
\end{abstract}

\section*{Introduction}
\begin{figure}[t] 
\centerline{\psfig{file=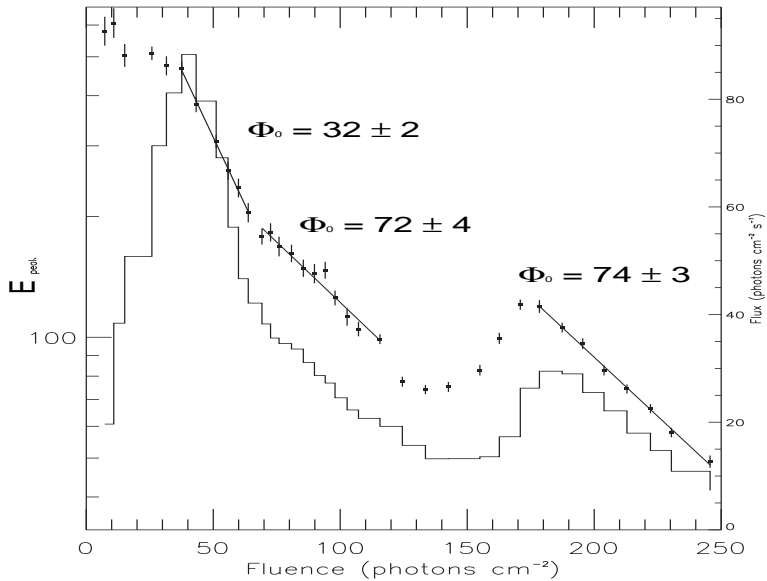,height=3.0in,width=4in}}
\vspace{10pt}
\caption{Evolution of the spectral energy break (squares) and photon flux
(histogram) as a function of
integrated photon flux $\Phi$ for GRB 921207 as seen by the BATSE LAD 0. 
The cooling decay constant $\Phi_{0}$ is shown for the three observed
pulses. The second and third pulses are examples of the
observed ``invariance'' of cooling rate seen within gamma-ray bursts.} 
\label{fig1}
\end{figure}

Much progress has been made in the past few years regarding the spectral
evolution of gamma-ray bursts.  Early studies described the time
evolution of a single hardness parameter in typically fewer than 20 bursts
[1-3].
Use of the BATSE LAD data has allowed time-resolved spectroscopy of
approximately one hundred bursts [4].
It has become clear that the time-integrated spectra is not
at all representative of the time-resolved spectra [5,6].
Several new trends have also
been found in analyses of the time-resolved spectra.  In particular, Liang \&
Kargatis [7] discovered a unique relationship between $\rm{E_{peak}}$ (the
energy at which the $\nu F_{\nu}$ spectrum is a maximum) and the photon fluence 
$\rm{\Phi(t) \equiv \int_{t'=0}^{t'=t} F_{N}(t') dt'}$, where
\begin{equation}
\rm{E_{peak}} = \rm{E_{peak}}^{(0)}~e^{-\frac{\Phi}{\Phi_{0}}}
\end{equation}
and $\Phi_{0}$ represents an effective cooling constant.  It was also
reported 
that $\Phi_{0}$ appeared to be invariant from pulse to pulse within some
multipulse bursts [7]. 
Figure \ref{fig1} shows an example of this observed invariance in cooling
rate for GRB 921207 as seen by BATSE LAD 0.  

If the decay of $\rm{E_{peak}}$ with respect to 
$\Phi$ is assumed to be the radiative cooling of a plasma, then energy
conservation dictates a proportionality between $\Phi_{0}$ and the number
of cooling leptons in the plasma [7]. 
This in turn suggests that multiple pulses
observed within a single burst originate from the same plasma. 
Verification of the observed ``cooling rate invariance'' would have
serious implications on gamma-ray burst models. 
A scenario in which a single plasma emits several pulses of radiation 
is difficult to reconcile with many catastrophic burster models
[8].  
As a strict test of the conclusions of Liang \& Kargatis [7] 
we determine the probability that the degree of invariance that they observed
in their bursts would originate from bursts with no intrinsic invariance at all.
As described below, this probability is high enough to raise doubts about 
any intrinsic invariance.

\section{Testing the Invariance of Cooling Rates within Bursts}
Liang \& Kargatis [7] fitted exponential decays of
$\rm{E_{peak}}$ with respect to fluence $\Phi$ in multipulse bursts and reported
that the decay constant $\Phi_0$ 
did not change from pulse to pulse in many of their bursts.
We tested this conclusion by randomly sampling values from the observed
distribution of $\Phi_0$.  
Examining the first pulse of 57 bursts, we found 
that we can approximate this distribution as
$\Phi_0 = 10^{1.53\pm0.39}$ ergs $\rm{cm}^{-2}$.   We also assigned a
$1\sigma$-confidence region
for each random value of $\Phi_0$ based on the observed
distribution $\sigma_{\Phi_0} / {\Phi_0} = 10^{-0.76\pm0.69}$.

In selecting the $\rm{E_{peak}}$ decays manually, it is likely that 
some systematic bias occurs
favoring the trends we are looking for, namely well fit decays with similar
decay rates.  To account for this, we used two different methods for
selecting
pulse decays.
For the case we designate ``Best'', we selected
the decay phases of pulses by directly examining plots of $\rm{E_{peak}}$ versus
$\Phi$.
For the case we
designate ``Worst'', we instead selected decay phases by examining plots of
photon flux versus time.   In both cases, however, we fit out exponential
decay to $\rm{E_{peak}}$ versus $\Phi$.  Thus, in the ``Worst'' case scenario,
we have blindly selected what portions of the $\rm{E_{peak}}-\Phi$ plot will be
fit.
Occasionally, there
are some pulses where $\rm{E_{peak}}$ may actually be rising with respect to
$\Phi_{0}$ in what was chosen to be a decay phase.  We discard
these instances of ``negative decay''.   What is left serves as an
\emph{extreme} limit of what our final results would have been
without our systematic bias.

We compared two of each bursts' M pulses at a time using
\begin{equation}
        \rm{X}_{ij}^{2} = \frac{[\Phi_{0}(i) - \Phi_{0}(j)]^{2}}
                          {\sigma_{\Phi_{0}(i)}^{2} +
\sigma_{\Phi_{0}(j)}^{2}}
\end{equation}
and then distilled the comparisons within each burst into a single
statistic to represent that burst (See Table \ref{table1} for definitions).
Each of these statistics are tailored for different
null hypotheses.
The statistic $G_{1}$ tests if \emph{at least} two pulses
in a burst are similar (and thus ``invariant''), while $G_{2}$ tests 
if \emph{all} the pulses
have a similar decay constant.  $G_{3}$
tests for either a single good pairing or
several moderately close pairings.  We believe that this last statistic 
is the most reasonable for testing our results since it does not 
require that \emph{all} pulses decay at the same rate (as $G_{2}$ does) but
also does not discard information about multiple pulses repeating (as $G_{1}$ does).
Finally, we calculated a table of probabilities P$(G,\rm{M})$
for our goodness-of-fit statistics $G$ based on random sampling
from our distributions of $\Phi_{0}$ and $\sigma_{\Phi_0} / {\Phi_0}$.
\begin{table}
\caption{The probabilities of getting such good values for each of our
three goodness-of-fit statistics $G$, where the ``Best'' and ``Worst'' 
case selections are discussed in the test.
Any reasonable analysis should fall somewhere between these two cases.
These probabilities are high enough $( > 0.001)$ to suggest that any
invariance seen in the data is purely coincidental.}
\vspace{10pt}
\label{table1}
\begin{tabular}{lll}
& ``Best'' Selection & ``Worst'' Selection \\
\hline
\hline
\\
P($G_{1} = \rm{min}~X_{ij}^{2}$) & 0.065 &  0.26 \\
\\
P($G_{2} = \rm{\sum_{i=1}^{M-1}\sum_{j=i+1}^{M}X_{ij}^{2}}$) & 0.0014 & 0.13 \\
\\
P($G_{3} = \rm{\prod_{i=1}^{M-1}\prod_{j=i+1}^{M}X_{ij}^{2}}$) &  0.012 & 0.12  \\ 
\\
\end{tabular}
\end{table}
We define P$(G,\rm{M})$ to be the probability of randomly getting a value of $G$
or lower if we compare $\rm{M}$ pulses drawn randomly from the global distribution
of $\Phi_0$.  Examining the distibution of P values
for our sample of 40 ``Best''
multipulse bursts, we found K-S probabilities of $0.1-6.5\%$ that
the observed ``repetitions'' occur by chance [9].
This number is much 
higher ($12-26\%$) if we remove our selection systematics.  These high
probabilities suggest that the observed invariance in cooling rate is not 
physical.

\section{Discussion}

An invariant cooling rate for pulses in multipulse gamma-ray bursts is 
certainly an exciting possibility.  Proof of such a pattern would
place hard limits on burst emission mechanisms.
Liang and Kargatis [7] found that 7 out of the 12 multipulse bursts 
in their sample were
consistent with an invariant decay (difference $< 1\sigma$).
However, using our
``Best'' selection and the $G_{1}$ goodness-of-fit measure,
we expect to randomly get $5.7\pm1.3$ bursts out of 12 consistent 
with having an invariant decay.
We conclude that no intrinsic invariance of $\Phi_{0}$ is required to
explain the observations reported in Liang and Kargatis [7]. 
The observed
``repeating'' cooling rates are consistent with being randomly selected
from the observed distribution of $\Phi_{0}$ values. 
While these new results
do not necessarily rule out cooling rate invariance in some bursts, they do
show that the observations as an ensemble do not require such an invariance.

\vspace{0.25in}

\noindent AC thanks NASA-MSFC for his GSRP fellowship. This work is partially
supported by NASA grant NAG5-3824.


\begin{references}
\bibitem{gole83}Golenetskii, S.V. {\it et al.}, Nature\ {\bf 306}, 451 (1983).
\bibitem{norr86}Norris, J.P., {\it et al.}, Ap. J.\ {\bf 301}, 213 (1986).
\bibitem{band92}Band, D. {\it et al.}, AIP Conf. Proc. {\bf 265}, 169
(1992).
\bibitem{pree98}Preece, R. {\it et al.}, these proceedings, (1998). 
\bibitem{crid97}Crider, A., {\it et al.}, Ap. J.\ {\bf 479}, L39 (1997).
\bibitem{crid98}Crider, A., Liang, E.P, \& Preece, R. D., these proceedings,
(1998).
\bibitem{lian96}Liang E., and Kargatis, V., {\it Nature}\ {\bf 381}, 49 (1996).
\bibitem{mesz93}M\'{e}sz\'{a}ros, P. \& Rees, M., Ap. J.\ {\bf 405}, 278 (1993).
\bibitem{pres92}Press, W.H., Teukolsky, S.A, Vetterling, W.T., \& Flannery, B.P.,
Numerical Recipes in C, 2nd ed, Cambridge: Cambridge Univ. Press\ (1992).
\end{references}
\end{document}